\documentclass[letterpaper, 10 pt, conference]{ieeeconf} 
\IEEEoverridecommandlockouts                              

\title{}
\author{}
\let\labelindent\relax
\usepackage{enumitem}
\usepackage{iftex}
\ifPDFTeX
\usepackage[utf8]{inputenc}
\fi

\usepackage[english]{babel}
\usepackage[autostyle]{csquotes}
\usepackage{listings}

\usepackage{ifthen} 
\usepackage{xparse}

\usepackage{tabularx}   
\usepackage{booktabs}   

\usepackage{mathtools}  
\usepackage{amssymb}    
\usepackage{siunitx}   

\usepackage{bm}

\usepackage{caption}
\usepackage{subcaption}

\usepackage{textcmds}

\usepackage[noend]{algorithm,algorithmic}

\usepackage{nccmath}

\usepackage{graphicx} 
\usepackage{caption}
\usepackage{subcaption}
\usepackage{cite}
\usepackage{pgfplots}
\usepackage{tikz}
\usepgfplotslibrary{external}
\usetikzlibrary{shapes}
\usetikzlibrary{positioning}
\usetikzlibrary{arrows,decorations.markings}
\usepgfplotslibrary{fillbetween}

\DeclareUnicodeCharacter{221E}{$_\infty$}

\usepackage{varwidth}
\usepackage{subcaption}
\usepackage[hidelinks]{hyperref}

\title{Towards remote fault detection by analyzing communication priorities}

\author{Alexander Gr\"afe$^{1}$, Dominik Baumann$^{2}$, and Sebastian Trimpe$^{1}$ 
\thanks{This work was supported in part by the German Research Foundation (DFG) within the priority program 1914 (grants TR 1433/1-1 and TR 1433/1-2), and the Federal Ministry of Education and Research (BMBF) and
the Ministry of Culture and Science of the German State of North Rhine-Westphalia
(MKW) under the Excellence Strategy of the Federal Government and the Länder.} \thanks{$^{1}$
		Institute for Data Science in Mechanical Engineering, RWTH Aachen University, 52068 Aachen, Germany. Email: \{alexander.graefe, sebastian.trimpe\}@dsme.rwth-aachen.de
	}
\thanks{$^{2}$ Department of Information Technology, Uppsala University, Sweden. Email: dominik.baumann@it.uu.se}
}

\newtheorem{theorem}{Theorem}

\begin{document}
	{\onecolumn \begin{center} Accepted for publication at the proceedings of the 61st IEEE Conference on Decision and Control\end{center}
		
		\noindent\fbox{%
			\parbox{\textwidth}{%
				© 2022 IEEE. Personal use of this material is permitted. Permission from IEEE must be obtained for all other uses, in any current or future media, including reprinting/republishing this material for advertising or promotional purposes, creating new collective works, for resale or redistribution to servers or lists, or reuse of any copyrighted component of this work in other works.
			}%
		}
	}
\twocolumn
		\newpage
	\maketitle
	\begin{abstract}
		The ability to detect faults is an important safety feature for event-based multi-agent systems. In most existing algorithms, each agent tries to detect faults by checking its own behavior. But what if one agent becomes unable to recognize misbehavior, for example due to failure in its onboard fault detection?  To improve resilience and avoid propagation of individual errors to the multi-agent system, agents should check each other remotely for malfunction or misbehavior.  In this paper, we build upon a recently proposed predictive triggering architecture that involves communication priorities shared throughout the network to manage limited bandwidth.  We propose a fault detection method that uses these priorities to detect errors in other agents.  The resulting algorithms is not only able to detect faults, but can also run on a low-power microcontroller in real-time, as we demonstrate in hardware experiments.
	\end{abstract}
	
	\newcommand{\fakepar}[1]{\vspace{1mm}\noindent\textit{#1.}}
\newcommand{\capt}[1]{\mdseries{\emph{#1}}}
\newcommand*{\matlab}{\textit{Matlab}}
\newcommand*{\simulink}{\textit{Simulink}}
\newcommand*{\purepursuit}{\textit{Pure Pursuit}}
\newcommand*{\carCup}[0]{\textit{Carolo-Cup}~}
\newcommand*{\ros}[0]{\textit{ROS2}~}
\newcommand*{\psaf}[0]{Projektseminar Autonomes Fahren~}
\newcommand*{\darpa}[0]{\textit{DARPA}~}
\newcommand*{\opencv}[1]{\textit{OpenCV}}

\newcommand*{\jt}{\textit{Jetson TX2}}
\newcommand*{\jpi}{\textit{SDK Manager}}

\newcommand*{\phil}{\ensuremath{\varphi_\textrm{L}}}
\newcommand*{\philmax}{\ensuremath{\varphi_\textrm{L,max}}}
\newcommand*{\dphilmax}{\ensuremath{\dot{\varphi_\textrm{L,max}}}}

\newcommand*{\mat}[1]{{\ensuremath{\mathrm{\textbf{#1}}}}}
\newcommand*{\ma}[1]{{\ensuremath{\boldsymbol{\mathrm{#1}}}}}
\newcommand*{\mas}[1]{\ensuremath{\boldsymbol{#1}}}
\newcommand*{\ve}[1]{{\ensuremath{\boldsymbol{#1}}}}
\newcommand*{\ves}[1]{\ensuremath{\boldsymbol{\mathrm{#1}}}}

\newcommand*{\AP}{\ensuremath{\mathrm{AP}}}
\newcommand*{\doti}{\ensuremath{(i)^\cdot}}

\newcommand*{\inprod}[2]{\ensuremath{\langle #1,\,#2 \rangle}}


\newcommand*{\ud}{\ensuremath{\mathrm{d}}}

\newcommand*{\tn}[1]{\textnormal{#1}}

\newcommand*{\mrm}[1]{\ensuremath{\mathrm{#1}}}

\newcommand*{\transp}{\ensuremath{\mathrm{T}}}

\newcommand*{\rang}{\ensuremath{\operatorname{rg}}}

\newcommand*{\grpsb}[2]{\ensuremath{\left(#1\right)_{#2}}}
\newcommand*{\grprsb}[2]{\ensuremath{\left(#1\right)_{\mathrm{#2}}}}

\newcommand*{\normd}[2]{\ensuremath{\frac{\mathrm{d}#1}{\mathrm{d}#2}}}
\newcommand*{\normdat}[3]{\ensuremath{\left.\frac{\mathrm{d} #1}{\mathrm{d} #2}\right|_{#3}}}

\newcommand*{\matd}[2]{\ensuremath{\frac{\mathrm{D} #1}{\mathrm{D} #2}}}
\newcommand*{\matdat}[3]{\ensuremath{\left.\frac{\mathrm{D} #1}{\mathrm{D} #2}\right|_{#3}}}

\newcommand*{\partiald}[2]{\ensuremath{\frac{\partial #1}{\partial #2}}}
\newcommand*{\partialdat}[3]{\ensuremath{\left.\frac{\partial #1}{\partial #2}\right|_{#3}}}

\newcommand*{\FT}[1]{\ensuremath{\mathfrak{F}\left\{#1\right\}}}
\newcommand*{\FTabs}[1]{\ensuremath{\left|\mathfrak{F}\left\{#1\right\}\right|}}
\newcommand*{\IFT}[1]{\ensuremath{\mathfrak{F}^{-1}\left\{#1\right\}}}
\newcommand*{\DFT}[1]{\ensuremath{\mathrm{DFT}\left\{#1\right\}}}
\newcommand*{\DFTabs}[1]{\ensuremath{\left|\mathrm{DFT}\left\{#1\right\}\right|}}
\newcommand*{\Laplace}[1]{\ensuremath{\mathfrak{L}\left(#1\right)}}
\newcommand*{\InvLaplace}[1]{\ensuremath{\mathfrak{L^{-1}}\left(#1\right)}}
\newcommand*{\invtrans}{\ensuremath{\quad\bullet\!\!-\!\!\!-\!\!\circ\quad}}
\newcommand*{\trans}{\ensuremath{\quad\circ\!\!-\!\!\!-\!\!\bullet\quad}}

\newcommand*{\textcompstdfont}[1]{{\fontfamily{cmr} \fontseries{m} \fontshape{n} \selectfont #1}}

\newcommand*{\mlfct}[1]{\texttt{#1}}

\newcommand*{\UL}[2]{#1_\mathrm{#2}}
\newcommand*{\ULi}[2]{#1_{#2}}
\newcommand*{\dy}[0]{\dot{y}}
\newcommand*{\ddy}[0]{\ddot{y}}
\newcommand*{\receivedPower}[0]{\frac{A}{y^2+h^2}}
\newcommand*{\dD}[0]{\dot{D}}
\newcommand*{\tf}[0]{\UL{t}{f}}
\newcommand*{\umax}[0]{\UL{u}{max}}

\newcommand*{\x}[1]{\UL{x}{#1}}
\newcommand*{\xv}[0]{\ve{x}}
\newcommand*{\la}[1]{\UL{\lambda}{#1}}
\newcommand*{\ua}[0]{\UL{u}{+}}
\newcommand*{\ub}[0]{\UL{u}{-}}
\newcommand*{\Ht}[1]{\tilde{H}(#1)}
\newcommand*{\Dp}[0]{\UL{D}{p}}
\newcommand*{\Dm}[0]{\overline{\dot{D}}}
\newcommand*{\fdynamic}[0]{\UL{\ve{f}}{dynamic}(\ve{y}, u)}
\newcommand*{\fE}[0]{\UL{f}{E}(\ve{y}, u)}
\newcommand*{\fD}[0]{\UL{f}{D}(\ve{y}, u)}
\newcommand*{\vl}[0]{\UL{v}{l}}
\newcommand{\vex}[0]{\ve{x}}
\renewcommand*{\of}[1]{\left(#1\right)}
\newcommand*{\vexof}[1]{\vex\of{#1}}
\newcommand*{\vexul}[1]{\UL{\vex}{#1}}
\newcommand*{\vexulof}[2]{\UL{\vex}{#1}\of{#2}}
\newcommand*{\vexuli}[1]{\ULi{\vex}{#1}}
\newcommand*{\hatvexuli}[1]{\ULi{\hat{\vex}}{#1}}
\newcommand*{\vexuliof}[2]{\ULi{\vex}{#1}\of{#2}}
\newcommand*{\hatvexuliof}[2]{\ULi{\hat{\vex}}{#1}\of{#2}}
\newcommand*{\pcom}[0]{{P\ul{com}}}
\newcommand*{\pcomhat}[0]{\hat{P}\ul{com}{}}
\newcommand{\weight}[2]{w\uliof{#1}{#2}}
\newcommand{\history}[0]{\mathcal{E}}
\newcommand{\weighthat}[2]{\hat{w}\uliof{#1}{#2}}

\newcommand*{\veerr}[0]{\ve{e}}
\renewcommand*{\ul}[1]{_{\mathrm{#1}} }
\newcommand*{\uliof}[2]{_{#1}{\of{#2}}}
\newcommand*{\uli}[1]{_{#1} }
\newcommand*{\koop}[0]{\mathcal{K}}
\newcommand*{\koopt}[0]{\mathcal{K}^t}
\newcommand*{\gof}[1]{g\of{#1}}
\newcommand*{\gul}[1]{\ULi{g}{#1}}
\newcommand*{\gofxof}[1]{\gof{\vexof{#1}}}
\newcommand*{\phivexof}[1]{\phi\of{\vexof{#1}}}
\newcommand*{\veg}[0]{\ve{g}}
\newcommand*{\vegof}[1]{\ve{g}\of{#1}}
\newcommand*{\vegulof}[2]{\ve{g}_{\mathrm{#1}}\of{#2}}
\newcommand*{\vey}[0]{\ve{y}}
\newcommand*{\veyof}[1]{\vey\of{#1}}
\newcommand*{\veyul}[1]{\UL{\vey}{#1}}
\newcommand*{\tzero}[0]{\UL{t}{0}}
\newcommand*{\vepsi}[0]{\ve{\psi}}
\newcommand*{\vephi}[0]{\ve{\phi}}
\newcommand*{\veu}[0]{\ve{u}}
\newcommand*{\veuof}[1]{\veu\of{#1}}
\newcommand*{\veuul}[1]{\UL{\veu}{#1}}
\newcommand*{\vev}[0]{\ve{v}}
\newcommand*{\vevof}[1]{\vev\of{#1}}
\newcommand*{\vevul}[1]{\UL{\vev}{#1}}
\newcommand*{\xul}[1]{\ULi{x}{#1}}
\newcommand*{\aul}[1]{\UL{a}{#1}}
\newcommand*{\kul}[1]{\UL{k}{#1}}
\newcommand*{\uul}[1]{\ULi{u}{#1}}

\newcommand*{\maAuli}[1]{\ULi{\ma{A}}{#1}}
\newcommand*{\veeps}[0]{\ve{\epsilon}}
\newcommand*{\twonorm}[1]{||#1||_\mathrm{2}}

\newcommand*{\lambdamaxof}[1]{\UL{\lambda}{max}\of{#1}}
\newcommand*{\lambdaminof}[1]{\UL{\lambda}{min}\of{#1}}
\newcommand*{\B}[0]{Method B\:}
\newtheorem{satz}{Satz}[section]
\newtheorem{assumtion}{Annahme}[section]

\newcommand*{\tensorflowprob}{\textit{Tensorflow-Probability}~}

\newcommand*{\EX}[0]{\mathbb{E}}
\newcommand*{\tr}[0]{\mathrm{Trace}}
\newcommand*{\var }[0]{\mathrm{Var}}
\newcommand*{\given}[0]{\;\middle|\;}
\newcommand*{\gpess}[0]{g\ul{pess}{}}
\newcommand*{\gopt}[0]{g\ul{opt}{}}
\newcommand*{\tveerr}[0]{\tilde{\veerr}}
\newcommand*{\ofs}[1]{\left[#1\right]}

\DeclareDocumentCommand\e{ m g }{%
	{\veerr\uli{#1}%
		\IfNoValueF {#2} {\of{#2}}%
	}%
}

\DeclareDocumentCommand\prio{ m g }{%
	{g\uli{#1}%
		\IfNoValueF {#2} {\of{#2}}%
	}%
}

\newcommand*{\matA}[1]{\ma{A}\uli{#1}}
\newcommand*{\matGamma}[0]{\ma{\Gamma}}
\newcommand*{\matGammaOpt}[0]{\ma{\Gamma}\ul{opt}}

\newcommand*{\matQ}[1]{{\ma{Q}\uli{#1}}}
\newcommand*{\matP}[1]{\ma{P}\uli{#1}}
\newcommand*{\eps}[2]{\epsilon\uli{#1}\of{#2}}
\newcommand*{\alp}[2]{\alpha\uli{#1}\of{#2}}
\newcommand*{\pfail}[0]{P\ul{loss}}
\newcommand{\priority}[2]{g\uli{#1}\of{\errorest{#1#1}{#2}}}
\newcommand*{\diag}[0]{\text{diag}}
\newcommand*{\tp}[1]{\of{#1}}
\newcommand*{\state}[2]{\vex\uli{#1}\tp{#2}}
\newcommand*{\stateest}[2]{\hat{\vex}\uli{#1}\tp{#2}}
\newcommand*{\noise}[2]{\ve{w}\uli{#1}\tp{#2}}
\newcommand*{\processnoise}[2]{\ve{v}\uli{#1}\tp{#2}}
\newcommand*{\errart}[2]{\ve{\tilde{e}}\uli{#1}\tp{#2}}
\newcommand*{\errorest}[2]{\ve{e}\uli{#1}\tp{#2}}
\newcommand*{\barerrorest}[2]{\ve{\bar{e}}\uli{#1}\tp{#2}}
\newcommand*{\errorestmeas}[2]{\ve{e}\uli{#1, meas}\tp{#2}}
\newcommand*{\erroragents}[2]{\hat{\ve{e}}\uli{#1}\tp{#2}}
\newcommand*{\erroragentstotal}[2]{\underline{\hat{\ve{e}}}\uli{#1}\tp{#2}}
\newcommand*{\erroresttotal}[2]{\underline{\ve{e}}\uliof{#1}{#2}}
\newcommand*{\matB}[1]{\ma{B}\uli{#1}}
\newcommand*{\matV}[1]{\ma{V}\uli{#1}}
\newcommand*{\matVti}[1]{\ma{\tilde{V}}\uli{#1}}
\newcommand*{\matW}[1]{\ma{W}\uli{#1}}
\newcommand*{\matAT}[1]{\left(\ma{A}\uli{#1}^\transp\right)}
\newcommand*{\matF}[1]{\ma{F}\uli{#1}}
\newcommand*{\rec}[2]{\gamma\uli{#1}\tp{#2}}
\newcommand*{\Cov}{\mathrm{Cov}}
\newcommand*{\matAti}[1]{\tilde{\ma{A}}\uli{#1}}

\newcommand{\io}[1]{\Gamma\of{#1}}
\newcommand{\oi}[1]{\Lambda\of{#1}}

\newcommand{\matAd}[1]{\tilde{\mat{A}}\ul{d, \mathrm{#1}}}
\newcommand{\ri}[1]{{r\uli{#1}}}
\newcommand{\di}[1]{{d\uli{#1}}}
\newcommand{\nii}[1]{{n\uli{#1}}}
\newcommand{\deltai}[1]{\Delta\uli{#1}}
\newcommand{\betai}[1]{\beta\uli{#1}}
\newcommand{\alphai}[1]{\alpha\uli{#1}}
\newcommand{\zetai}[1]{\zeta\uli{#1}}
\newcommand{\lmax}[1]{\lambda\ul{max}\of{#1}}
\newcommand{\lmin}[1]{\lambda\ul{min}\of{#1}}
\newcommand{\unknown}[0]{\left(\cdot\right)}
\newcommand*{\kt}[0]{\tilde{k}}
\newcommand*{\nz}[0]{\ve{n}\ul{0}}
\newcommand*{\norm}[2]{||#1||_{#2}}
\newcommand*{\diff}[0]{\text{d}}

\newcommand{\dalpha}[0]{\delta\alpha}
\newcommand{\setS}[0]{\mathcal{S}}
\newcommand{\cone}[0]{c\ul{1}\of{\setS}}
\newcommand{\ctwo}[0]{c\ul{2}\of{\setS}}
\newcommand{\vef}[0]{\ve{f}}
\newcommand{\dist}[0]{\mathcal{V}}
\newcommand{\measurement}[2]{\ve{x}\uliof{#1}{#2}}
\newcommand{\measurementnoise}[2]{\ve{w}\uliof{#1}{#2}}
\newcommand{\contr}[0]{\ve{c}}
	
	\section{Introduction}
In networked multi-agent systems, multiple agents, such as mobile robots in a smart factory~\cite{Baumann_2021} or autonomous cars~\cite{Yaqoob2020}, interact to jointly solve a task (e.g., to manufacture specific products or increase traffic throughput).
In such scenarios where autonomous agents act in the real world, safety is paramount.
To ensure safety, a key system requirement is the ability to detect hardware and software faults.
A first step towards improving reliability is ensuring that each agent constantly checks its own sensor readings. This is typically already considered in the area of fault detection \cite{Isermann_2005}. However, sole reliance on this on-board fault detection (FD) is hazardous. An agent may become unable to detect its own incorrect behavior. For example, a software bug or a malicious attack on the agent could lead to malfunction of the on-board FD, meaning the failure would go undetected by the local agent alone. To enhance resilience in multi-agent systems, each agent should therefore not only check its own sensor readings, but also try to detect potential faults in other agents by checking data received from them over the network. 

In this work, we consider remote FD for event-triggered multi-agent systems with heterogeneous nonlinear dynamics. When multiple systems use the same network to transmit data, communication becomes a scarce resource due to limited bandwidth. This situation has rendered event-triggered communication increasingly popular for networked systems in the last two decades \cite{NCSET}. However, this form of communication comes with challenges for remote FD. Messages from a particular agent will not always be available. For example, if the event-triggered communication scheme prevents messages from being sent after a fault occurs, faults may be detected too late or missed altogether. 
Instead of building on such sparse event-triggered signals, we propose an FD method that uses communication priorities. These are low-dimensional network commands that are periodically shared across the network to manage network traffic.

\fakepar{Main idea} We build upon a recently proposed event-triggered control structure called predictive triggering \cite{Baumann_2019, MASTRANGELO2019, PredictiveTriggering}. It takes limited communication resources into account by allowing only a subset of agents to communicate in every communication round. This is achieved by deriving a priority for each agent, which is subsequently shared. Only agents with the highest priority are allowed to send. 

Although the messages that agents exchange, e.g., state measurements, may be high-dimensional and thus costly to transmit over the network, the low-dimensional priority of the agent is available periodically. The priorities are therefore predestined to be used to monitor the agent's condition. While priorities have been introduced in predictive triggering to quickly adapt limited communication resources to current need, we will show in this paper how they can also be used to simultaneously detect faults remotely.

This paper makes the following contributions:
\begin{enumerate}[wide, labelindent=0pt]
	\item A novel fault detection algorithm for networked multi-agent systems based on low-dimensional priority signals;
	\item Statistical methods for bounding false positives for two versions of the algorithm: with (i) a static detection threshold and (ii) a time variable threshold that is adaptive to the network load;
	\item 
	Efficient implementation with low computational footprint, allowing for implementation on low-power devices;
	\item Experimental evaluation on multiple cart-pole systems interconnected through a multi-hop wireless network.
\end{enumerate}

\section{Related work}

Fault detection, in general, is a broad field. An overview of classical FD algorithms can be found in \cite{Isermann_2005} and \cite{WILHELM2021}.
In this work, we focus on detecting faults in networked multi-agent systems. For this scenario, the key challenges are delayed and missing measurements due to network imperfections and limited communication resources. 

Some existing studies deal with the detection of faults based on measurements received over a communication network. The works in \cite{Zhang_2004, Zhang2007, Wang_2009, Wang_2016} focus on networked control systems with missing measurements caused by data dropouts. The approach in \cite{Wang_2009}, for example, adapts the decision threshold of the FD based on the communication history, an approach we will also adopt. However, these works do not deal with event-triggered systems,  the focus of our paper.

As mentioned above, event triggering poses a challenge for remote FD. Wang et al.\ \cite{Wang_2017} design an FD for event-triggered control using a Luenberger observer to generate residuals for the FD even if no new measurement is received.
However, this approach only incorporates new information when new measurements arrive over the network. If this rarely happens, faults might be detected too late.
Instead, we base our FD not on measurements but on low-dimensional communication priorities that can be sent periodically at low cost.

Solowjow et al.\ \cite{Solowjow2_2018} use the communication signal in event-triggered control to detect changes in system behavior.
Although some of the techniques described here bear similarities to their work (esp. statistical tests), the objective, setting, and proposed methods differ. While \cite{Solowjow2_2018} aims to trigger model learning for a simpler event-triggered scheme that does not consider explicit bandwidth constraints, we develop FD for predictive triggering that accounts for such constraints. Importantly, we do not build our detector on the event-triggered signal, but on communication priorities.

\section{Problem setting}
The setting in this work includes $N$ agents that must solve a task cooperatively, e.g., to stabilize and synchronize with each other as in \cite{NetworkControlGoesWireless, PredictiveTriggering}. These agents are physically independent, i.e., their open-loop dynamics are not coupled, and connected over a wireless network. This network has limited bandwidth, which does not allow all agents to transmit their data simultaneously. To decide which agents get to transmit and which control input to apply in each time step, we employ the predictive triggering scheme \cite{Baumann_2019, MASTRANGELO2019}, which has been demonstrated to yield effective bandwidth usage in experiments on a wireless multi-hop network \cite{PredictiveTriggering}.
\subsection{Networked control system architecture}
We briefly introduce the predictive triggering scheme as background to this work.
We model the agents as nonlinear stochastic systems
\begin{equation}
	\label{eq:systemdefinition}
	\state{i}{k+1} = \vef\uli{i}\of{\state{i}{k}, \ve{u}\uli{i}\tp{k}}+ \processnoise{i}{k},
\end{equation}
with $\state{i}{k}\in\mathbb{R}^\nii{i}$, $\ve{u}\uli{i}\tp{k}\in\mathbb{R}^{b\uli{i}}$, and $\vev\uliof{i}{k} \sim\mathcal{V}\uli{i}$ being independent and identically distributed (i.i.d.) process noise with distribution $\mathcal{V}\uli{i}$, which has zero mean and finite variance. We further assume that the full state $\measurement{i}{k}$ can be measured by its corresponding agent. 

To achieve a joint objective, agents can exchange information over the network (Figure \ref{fig:predictiveTriggering}). We consider a many-to-all communication protocol, e.g., Mixer \cite{Mixer, PredictiveTriggering}. The communication occurs in dedicated rounds. At each discrete timestep $k$, all $N$ agents receive new information sent by the sending $M$ agents at time $k-1$.
Due to limited communication resources, the network only allows $M\ll N$ agents to send their messages in one round. For scheduling, all agents send priorities during the communication round to all others. Then, the $M$ agents with the highest priorities transmit their message in the next round. A real-world implementation of the described network and scheduling technique can be found in\cite{PredictiveTriggering}, which has shown that the total bandwidth savings exceed the additional communication overhead of sending priorities.

\subsection{Event-triggered controller}
To achieve the distributed task, the controller of each agent is split into two parts. The first part generates the input signal $\veu\uli{i}\tp{k}$ for the agent, while the second part calculates the priority to trigger communication.
Each agent generates its input using a control strategy $\veu\uli{i}\tp{k} = \contr\uli{i}\of{\measurement{i}{k}, \stateest{i}{k} }$, where $\stateest{i}{k}=\{\stateest{\ell i}{k}|\forall \ell\in\{1,2,..,N\},~\ell\neq i\}$ is the set of extrapolates of the other agents' states. 
These extrapolates are calculated based on the knowledge of the system dynamics and data communicated over the network.
In particular, agent $i$'s extrapolate of agent $\ell$'s state is
\begin{equation}
	\label{eq:stateest}
	\begin{split}
		&\stateest{\ell i}{k+1} =\\ &\begin{cases}\vef\uli{\ell}\of{\stateest{\ell}{k}, \contr\uli{\ell}\of{\stateest{\ell i}{k}, \stateest{\ell}{k}}} & \text{if~}\rec{\ell}{k} = 0 \\
			\vef\uli{\ell}\of{\measurement{\ell}{k}, \contr\uli{\ell}\of{\measurement{\ell}{k}, \stateest{\ell}{k}}} & \text{if~}\rec{\ell}{k} = 1,
		\end{cases}
	\end{split}
\end{equation}
where $\rec{\ell}{k} = 1$ means that the state of agent $\ell$ is sent. In this case, the received measurement $\measurement{\ell}{k}$ is predicted one timestep into the future since communicating states over the network introduces a delay of one timestep. In case no update is received, the state estimate is simply extrapolated using the system dynamics model.

The second part of the controller is the priority calculation to trigger communication. In our setting, the agents do not compete for resources but cooperatively decide which agents need to communicate based on priorities. The priority of agent $i$ at time $k$ is called $g\uli{i}(k)$. The $M$ agents with the highest priorities are allowed to send in the next communication round. Thus, the higher the agent's need to communicate, the higher its priority should be. The key concept behind the priority calculation is so-called predictive triggering \cite{Baumann_2019, MASTRANGELO2019, PredictiveTriggering}.

If past disturbances $\processnoise{i}{k}$ were very low, the extrapolation in (\ref{eq:stateest}) would lead to an accurate state estimation, and there would be no need to communicate. If the system exhibits a stronger disturbance, the estimation error becomes large. In this case, the estimated state should be updated by communicating the current measurement over the network. Hence, the estimation error is directly related to the need to communicate and can be used to calculate the priority. 
All agents are aware that the other agents are trying to estimate their state. They calculate the same estimate and compare it with their actual states.
That is, agent $i$ is able to calculate the estimation error $\errorest{ii}{k} = \measurement{i}{k}-\stateest{ii}{k} = \measurement{i}{k}-\stateest{i\ell}{k}$. Various priority measures based on the error $\errorest{ii}{k}$ may be conceivable. In the following, we develop our algorithm for a generic priority measure for which the priorities are higher if the error is larger. We make our specific choice of priority measure concrete when we present our results in Section \ref{sec:experiments}.

The presented scheme has been considered in \cite{PredictiveTriggering} which has proofed its stability for linear systems. Here, we assume that the controller yields a stable system and focus on FD.

\subsection{Problem}
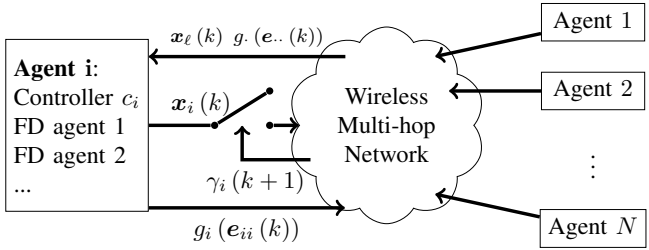
\begin{figure}
	\newcommand{\arrowwidth}{0.5mm}
	\resizebox{\columnwidth}{!}{%
		\begin{tikzpicture}
			\node [draw, minimum width=1.5cm, minimum height=2.5cm, align = left] (agent1) {\textbf{Agent $\bf{i}$}:\\Controller $c\uli{i}$\\ FD agent 1\\FD agent 2\\...};
			\node [cloud, draw,cloud puffs=10,cloud puff arc=120, aspect=1.0, inner ysep=1em, right of= agent1, xshift=3.5cm, align=center](network){Wireless\\Multi-hop\\Network};
			
			\node[circle,fill=white,inner sep=0pt,minimum size=0pt, left of=network, xshift=0.4cm] (networkleft) {};
			
			\draw[->, transform canvas={yshift=1cm}, line width=\arrowwidth] (networkleft.center) -- node [anchor=south] {$\medmath{\state{\ell}{k} \ \priority{\cdot}{k}}$}
			(agent1);
			\draw[->, transform canvas={yshift=-1.2cm}, line width=\arrowwidth] (agent1) -- node [anchor=north] {$\priority{i}{k}$} (networkleft);
			
			\node[circle,fill=black,inner sep=0pt,minimum size=3pt, right of=agent1, xshift=1cm] (a) {};
			
			\node[circle,fill=black,inner sep=0pt,minimum size=3pt, right of=a, xshift=-0.2cm, yshift=0.5cm] (a1) {};
			
			\node[circle,fill=white,inner sep=0pt,minimum size=0pt, right of=a, xshift=-0.6cm] (schalter) {};
			
			\node[circle,fill=white,inner sep=0pt,minimum size=0pt, below of=schalter, yshift=0.5cm] (schalterbelow) {};
			
			\node[circle,fill=white,inner sep=0pt,minimum size=0pt, right of=schalterbelow] (schalterbelowright) {};
			
			\node[circle,fill=black,inner sep=0pt,minimum size=3pt, right of=a, xshift=-0.2cm] (a2) {};
			
			\draw[-, line width=\arrowwidth] (agent1) -- node [anchor=south, pos=0.85] {$\state{i}{k}$} (a);
			
			\draw[-, line width=\arrowwidth] (a) -- (a1);
			\draw[->, line width=\arrowwidth] (a2) -- (network);
			
			\draw [->, line width=\arrowwidth] (schalterbelowright) -|  node [anchor=north, pos = 0.4] {$\rec{i}{k+1}$} (schalter);
			
			\node [draw, minimum width=1.5cm, minimum height=0.5cm, xshift=2cm, yshift=1.5cm, right of=network] (agent2) {Agent $1$};
			
			\node [draw, minimum width=1.5cm, minimum height=0.5cm, xshift=2cm, yshift=0.5cm, right of=network] (agent3) {Agent $2$};
			
			\node [, minimum width=1.5cm, minimum height=0.5cm, xshift=2cm, yshift=-0.5cm, right of=network] (agent4) {$\vdots$};
			
			\node [draw, minimum width=1.5cm, minimum height=0.5cm, xshift=2cm, yshift=-1.5cm, right of=network] (agent5) {Agent $N$};
			
			\node[circle,fill=white,inner sep=0pt,minimum size=0pt, right of=network, xshift=-0.3cm, yshift=1cm] (networkright) {};
			
			\draw[->, line width=\arrowwidth] (agent2) -- (networkright);
			
			\node[circle,fill=white,inner sep=0pt,minimum size=0pt, right of=network, xshift=-0.1cm, yshift=0.5cm] (networkright) {};
			
			\draw[->, line width=\arrowwidth] (agent3) -- (networkright);
			
				\node[circle,fill=white,inner sep=0pt,minimum size=0pt, right of=network, xshift=-0.3cm, yshift=-1cm] (networkright) {};
			
			\draw[->, line width=\arrowwidth] (agent5) -- (networkright);
			\node [below of=agent1, yshift=-1cm] (temp) {};
			\node [above of=agent1, yshift=1cm] (temp1) {};

		\end{tikzpicture}
	}
	\caption{Proposed event-triggered control system with FD. \capt{In every communication round, $M$ agents send their states to every agent in the network. Simultaneously, the network collects priorities $g$ of all $N$ agents. The agents with the highest priorities are allowed to send their states in the next round. Every agent has an FD to detect faults of other agents.}}
	\label{fig:predictiveTriggering}
\end{figure}

Our goal is to design a remote FD mechanism that allows agents to check each other for faults. 
This mechanism is shown in Figure \ref{fig:predictiveTriggering}: each agent runs $N$ fault detectors to check both itself \emph{and} all other agents for potential faults based on the priorities sent over the network.
If it detects a fault, it should take countermeasures, e.g. going into safe mode. This adds an additional security layer to the on-board FD. In this work, we focus on the FD only and do not discuss possible countermeasures.

The remote FD should capture the difference between measured and expected behavior. However, individual agents are not capable of directly measuring the states of other agents. They can only use data about those agents communicated over the network. While these states are rarely sent due to the triggering mechanism, the priorities are available in every communication round. Moreover, the priorities $\prio{i}{k}$ are functions of the difference between expected and measured behavior. We take advantage of these synergies and propose an FD that uses these priorities.

Since the system dynamics are stochastic, we employ a statistical test with the null hypothesis ``The system has no fault''.
If the null hypothesis is rejected, the FD algorithm triggers an alarm.
To improve efficiency, we want to ensure that the probability of false positives (faults detected by the algorithm although there is no fault in the system)
\begin{equation}
	\label{eq:firstordererror}
	P\of{\text{fault detected}\mid\text{no fault}} < \eta
\end{equation}
is bounded by $\eta\ll1$. The choice of $\eta$ is a trade-off between false positive and false negative rate.

We design two statistical tests. The first does not take the scheduling history into account, while the second does. This will lead to a more adaptive behavior of the second test.

	\section{Priority-based fault detection}
In this section, we present our main contribution and design two FD algorithms based on the communication priorities. Both algorithms define a decision threshold $\kappa\uli{i}$ to decide whether the output of the fault detector is \qq{Fault} or \qq{No Fault}. For both algorithms, we will show how to approximate $\kappa_i$ such that the probability of false positives is bounded by $\eta$. The first algorithm is straightforward and uses a static decision threshold as, e.g., in \cite{Solowjow2_2018, Wang_2017}. We call it: fault detector with static decision threshold (sFD).
 However, sFD ignores the communication history that could be used to construct a better bound. E.g., when the system has not been able to send for a longer time period, its current priority will most likely be larger. In this case, a static threshold will cause the false positive rate to rise. Hence, we also present an algorithm that includes a time-variant threshold, which is done in e.g. \cite{Wang_2009}. We call this algorithm: fault detector with dynamic decision threshold (dFD).
\subsection{Fault detector with static decision threshold}
We propose the following condition to check for a fault of an agent $i$ at time $k$
\begin{equation}
	\label{eq:faultdetectioncond}
	\sum_{\tau=0}^{d-1}g\uliof{i}{k-\tau} > \kappa\uli{i},
\end{equation}
where the horizon $d$ can be chosen freely. The sum acts as a low pass filter, avoiding fast oscillations in the detection decision. The larger $d$ is, the more stable the decision, but the longer the system needs to detect faults.
We seek to select $\kappa\uli{i}$ such that (\ref{eq:firstordererror}) is true. In our case, a false positive is generated if (\ref{eq:faultdetectioncond}) is true though the system has no faults. Thus, we require the probability that the sum of priorities exceeds the decision threshold given no fault, $P[\sum_{\tau=0}^{d-1}g\uliof{i}{k-\tau} > \kappa\uli{i}| \text{no fault}]$, to be lower than $\eta$.
Since the predictive triggering leads to a complicated probability distribution, 
this probability is analytically intractable. We therefore approximate $\kappa\uli{i}$ using Monte Carlo simulations (MC) of the multi-agent system. The system is simulated in its fault-free case, and the distribution of the sum of $d$ consecutive priorities is calculated by sampling. Then, $\kappa\uli{i}$ is selected such that it is equal to the $100(1-\eta)$th percentile. Since $\kappa\uli{i}$ is constant, this can be done before the deployment.

\subsection{Fault Detector with dynamic decision threshold}
Next, we show how the decision threshold can be adapted based on the communication history. In particular, the algorithm will change its decision threshold depending on when and how many times the agent communicated in the past. This comes with two principal challenges: \emph{(i)} as for the static case, the probability distribution is here also analytically intractable and must be approximated via MC, \emph{(ii)} if the threshold depends on the communication history, so too does the probability of false positives. Hence, not just one but many probabilities that grow exponentially (base two) with the length of the communication history must be approximated via MC. This becomes numerically intractable.

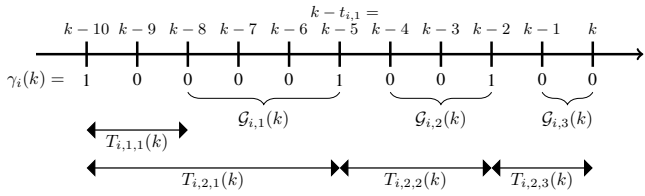
\begin{figure}
	\vspace{0.1cm}
	\newcommand{\drawwidth}{0.5mm}
	\resizebox{\columnwidth}{!}{%
		\begin{tikzpicture}
			\draw[->, line width=\drawwidth] (-2, 0) -- (10, 0);
			\draw[line width=\drawwidth] (-1, 0.25) -- (-1,-0.25);
			\node at (-1, 0.5){\small$k-10$};
			\node at (-1, -0.5){1};
			\draw[line width=\drawwidth] (0, 0.25) -- (0,-0.25);
			\node at (0, 0.5){\small$k-9$};
			\node at (0, -0.5){0};
			\draw[line width=\drawwidth] (1, 0.25) -- (1,-0.25);
			\node at (1, 0.5){\small$k-8$};
			\node at (1, -0.5){0};
			\draw[line width=\drawwidth] (2, 0.25) -- (2,-0.25);
			\node at (2, 0.5){\small$k-7$};
			\node at (2, -0.5){0};
			\draw[line width=\drawwidth] (3, 0.25) -- (3,-0.25);
			\node at (3, -0.5){0};
			\node at (3, 0.5){\small$k-6$};
			\draw[line width=\drawwidth] (4, 0.25) -- (4,-0.25);
			\node at (4, -0.5){1};
			\node at (4, 0.5){\small$k-5$};
			\node at (4.1, 0.8){\small $k-t\uli{i, \mathrm{1}}=$};
			\draw[line width=\drawwidth] (5, 0.25) -- (5,-0.25);
			\node at (5, -0.5){0};
			\node at (5, 0.5){\small$k-4$};
			\draw[line width=\drawwidth] (6, 0.25) -- (6,-0.25);
			\node at (6, -0.5){0};
			\node at (6, 0.5){\small$k-3$};
			\draw[line width=\drawwidth] (7, 0.25) -- (7,-0.25);
			\node at (7, -0.5){1};
			\node at (7, 0.5){\small$k-2$};
			\draw[line width=\drawwidth] (8, 0.25) -- (8,-0.25);
			\node at (8, -0.5){0};
			\node at (8, 0.5){\small$k-1$};
			\draw[line width=\drawwidth] (9, 0.25) -- (9,-0.25);
			\node at (9, 0.5){\small $k$};
			\node at (9, -0.5){0};
			
			\node at (-2, -0.5){$\gamma\uli{i}(k)=$};
			
			\draw[decorate,decoration={brace,amplitude=7pt,mirror}] (1, -0.75) -- (4, -0.75) node[midway, below,yshift=-7pt]{$\mathcal{G}\uli{i, \mathrm{1}}(k)$};
			
			\draw[decorate,decoration={brace,amplitude=7pt,mirror}] (5, -0.75) -- (7, -0.75) node[midway, below,yshift=-7pt]{$\mathcal{G}\uli{i, \mathrm{2}}(k)$};
			
			\draw[decorate,decoration={brace,amplitude=7pt,mirror}] (8, -0.75) -- (9, -0.75) node[midway, below,yshift=-7pt]{$\mathcal{G}\uli{i,\mathrm{3}}(k)$};
			
			\draw[line width=\drawwidth/2, triangle 90-triangle 90] (-1, -1.5) -- (1, -1.5);
			
			\node at (0, -1.75){$T\uli{i, \mathrm{1, 1}}(k)$};
			
			\draw[line width=\drawwidth/2, triangle 90-triangle 90] (-1, -2.25) -- (4, -2.25);
			
			\node at (1.5, -2.5){$T\uli{i, \mathrm{2, 1}}(k)$};
			
			\draw[line width=\drawwidth/2, triangle 90-triangle 90] (4, -2.25) -- (7, -2.25);
			
			\node at (5.5, -2.5){$T\uli{i, \mathrm{2, 2}}(k)$};
			
			\draw[line width=\drawwidth/2, triangle 90-triangle 90] (7, -2.25) -- (9, -2.25);
			
			\node at (8, -2.5){$T\uli{i, \mathrm{2, 3}}(k)$};
		\end{tikzpicture}
	}
	\caption{Splitting the scheduling history into smaller parts. \capt{The fault detection algorithm dFD splits the scheduling history into intervals, where the agent is not sending in between and is sending at the end. In this case $d=9$.}}
	\label{fig:faultdetectionalg}
\end{figure}
To overcome challenge (ii), the algorithm structures the past. In particular, the algorithm splits the time window $\left[k-d+1, k\right]$ with horizon $d$ into $H\uli{i}\of{k}$  periods $\mathcal{G}\uli{i, h}(k)$ ($h\in\{1,\cdots,H\uli{i}\of{k}\}$), with no communication in between (Figure \ref{fig:faultdetectionalg}). Hence, every period, except for the last, ends with a communication round. For each period, the fault detector sums up the priorities. If this sum is greater than a threshold
\newcommand{\sumprio}{s\uli{i}\of{k, \mathcal{G}\uli{i, h}(k)}}
\begin{equation}
	\label{eq:faultdetectionconditionVtwo}
	\sumprio\coloneqq\sum_{k-\tau\in\mathcal{G}\uli{i, h}(k)}g\uliof{i}{k-\tau} > \kappa\uli{i, h}\of{k},
\end{equation}
for at least one period $\mathcal{G}\uli{i, h}(k)$, a fault alarm is triggered.
The threshold $\kappa\uli{i, h}\of{k}$ is adapted based on the communication parameters of the corresponding period.

While most periods are fully characterized by a single parameter, the first and the last period play a special role.
The first period $\mathcal{G}\ul{i, 1}(k)$ ranges from time $k-d+1$ to time $k-t\uli{i, \mathrm{1}}$, where $t\uli{i, \mathrm{1}}$ is chosen such that $\rec{i}{k-t\uli{i, \mathrm{1}}}=1$ for the first time in this period. The period is parameterized by two different times, $T\uli{i, \mathrm{1, 1}}(k)$ and $T\uli{i, \mathrm{2, 1}}(k)$ (start and end delays). Both delays have three indices. The first, $i$, is the agent, the second indicates the start (one) or end (two) delay, and the third is the number of the period $\mathcal{G}\ul{i, h}(k)$ (in this case $h=1$). The start delay $T\uli{i, \mathrm{1, 1}}(k)$ represents the duration between the last communication time and the start of the period at $k-d+1$, and the end delay $T\uli{i, \mathrm{2, 1}}(k)$ the duration between the last communication time and the end of the period at time $k-t\ul{1}$. The delays can be obtained by the agent simply by looking at the past communication history. All following periods $\mathcal{G}\uli{i, h}(k)$ start directly after a communication occurred and are thus fully characterized by their length $T\uli{i,2,h}(k)$. Each of these periods, with the exception of the last, ends with a communication round. Thus, they all follow the same type of distribution parameterized by $T\uli{i,\mathrm{2}, h}(k)$. The last period ends at the current time $k$ and at this time the system may trigger or not. Therefore, the priorities in the last period follow a different distribution parameterized only by its length $T\uli{i, 2,H\uli{i}\of{k}}(k)$.
The periods $\mathcal{G}\uli{i, h}(k)$ in the middle ($1<h<H\uli{i}\of{k}$) are a special case of the first period, with start delay equal to one. To them, we thus add an additional parameter $T\uli{i, 1, h}=1$. This allows us to treat all periods for $h<H\uli{i}\of{k}$ in the same way. To simplify notation, we also define $T\uli{i, 1,H\uli{i}\of{k}} = 1$.

Given this partitioning, we can now set $\kappa\uli{i, h}\of{k} = \kappa\uli{i}(T\uli{i, 1, h}(k), T\uli{i, 2, h}(k), H\uli{i}\of{k}, a)$ in~\eqref{eq:faultdetectionconditionVtwo} to be dependent on the start and end delays, the number of periods $H\uli{i}\of{k}$, and whether the period is the last one ($a\in\left\{0, 1\right\}$). We wish to choose $\kappa\uli{i, h}\of{k}$ such that the probability of false positives is bounded by $\eta$. However, because of challenge (i), we need to perform MC simulations to approximate probabilities. This cannot be done in real time on a low-power microcontroller. We therefore propose to precalculate thresholds and save them in a lookup table, where the dimensions are the dependencies of the threshold. However, while $H\uli{i}\of{k}$ and $a$ are bounded ($H\uli{i}\of{k}\leq d$, $a\in\left\{0, 1\right\}$), this does not hold for $T\uli{i, 1, h}$ and $T\uli{i, 2, h}(k)$. We solve this by imposing the limits $T\uli{1, h}(k)\leq T\uli{i, 2, h}(k) \leq b$. The lookup table then has the size $b\times b\times d\times 2$, which is moderate for small $b,d$.
If $T\uli{i, 2, h}(k) > b$, we can conclude that the agent has not been allowed to communicate for longer than $b$ timesteps.
Its estimated state was always close to the real state, and the system behaved as expected in this period. Therefore, in this case we set $\kappa\uli{i}\of{T\uli{i, 1, h}(k), T\uli{i, 2, h}(k), h, H\of{k}} = \infty$, i.e., no fault will be triggered for this period. This significantly reduces the number of distributions to precalculate.

The thresholds in the lookup table for $T\uli{i, 2, h}(k) \leq b$ should be calculated such that the probability of false positives is bounded by $\eta$,
\begin{equation}
	\label{eq:firstordererrorVtwo}
	\begin{split} P\big[\bigcup\limits_{h=1}^{H\uli{i}\of{k}}\left(\sumprio > \kappa\uli{i, h}\of{k}\given \mathcal{G}\uli{i, h}(k), \text{nf}\right)\big]<\eta,
	\end{split}
\end{equation}
where we have abbreviated \qq{no fault} with \qq{nf}.
Note that the sets inside the probability are not probabilistically independent of each other. 
 Since only a fixed number of agents can communicate at a time, triggering decisions by one agent will affect the others. Consider the case where agent i last triggered at $k-\kappa$ while agent j did not trigger between $k-\kappa$ and $k-1$ but did at $k$. Then, we know that $\prio{j}(k-\kappa)<\prio{i}(k-\kappa)$ and $\prio{i}(k)<\prio{j}(k)$. Thus, since $\prio{j}(k)$ depends on $\prio{j}(k-\kappa)$, $\prio{j}(k)$ also depends on $\prio{i}(k-\kappa)$.

We trigger an alarm in the case where a single sum within the $H\uli{i}(k)$ periods exceeds the threshold. To prevent false positives, we
select $\kappa\uli{i}\of{T\uli{1, h}(k), T\uli{2, h}(k), h, H\of{k}}$ such that the probability of a false positive is bounded by $\eta/H\of{k}$.
We approximate these thresholds via MC as in the previous section.
If our approximation is exact, the first order error is bounded by $\eta$:
\begin{theorem}
	\label{th:firstorderboundVtwo}
	If for every $T\uli{i, 1, h}(k)\leq b$ the entries in the lookup table $\kappa\uli{i}\of{T\uli{i, 1, h}(k), T\uli{i, 2, h}(k), h, H\uli{i}\of{k}}$ fulfill
	\begin{equation}
		\label{eq:probfirstordereq}
P[\sumprio > \kappa\uli{i, h}\of{k}| \mathcal{G}\uli{i, h}(k),\text{nf}] < \frac{\eta}{H\uli{i}\of{k}},
	\end{equation}
	and $\kappa\uli{i}\of{T\uli{i, 1, h}, T\uli{i, 2, h}, h, H\uli{i}\of{k}} = \infty$ if $T\uli{i, 2, h}>b$,
	then (\ref{eq:firstordererrorVtwo}) holds true.
\end{theorem}
\begin{proof}
	We know that in cases where $T\ul{i, 2, h}>b$, the probability of a false positive is zero.
	Combining this and condition (\ref{eq:probfirstordereq}), we know that the left side probability in (\ref{eq:firstordererrorVtwo}) is smaller than the sum over all $P[\sumprio > \kappa\uli{i, h}\of{k}| \mathcal{G}\uli{i, h}(k), \text{nf}]$ \footnote{This can be recursively shown using $P(A\cup B) = P(A) + P(\bar{A}\cap B)\leq P(A)+P(B)$ for two random variables $A$ and $B$.}, which is smaller than $\eta$.
\end{proof}

The dFD algorithm adapts based on the communication history while maintaining guarantees on the probability of false positives. Note that the derivation in the proof shows that the method has some conservativeness. As it uses lookup tables combined with some logic to split the past, it requires few computational resources.

	\section{Experiments}
\label{sec:experiments}
To the best of our knowledge, there is currently no other method that detects faults based on communication priorities as used in predictive triggering. The aim of this section is thus not to compare the performance of our proposed methods with existing methods, but rather to show that communication priorities can be used to detect faults.
We demonstrate this in two experiments.
First, we test our FD algorithm in a simulation. This simulation allows us to generate many runs, enabling a quantitative evaluation of the algorithms. Second, we implement the FD algorithm on a cart-pole testbed with wireless communication \cite{CPSTestbed}. With these experiments, we show that our algorithm can run in real-time on a low-power microcontroller and detect faults of a real physical system. 
\subsection{Simulation results}
	\definecolor{grayyyy}{rgb}{0.8,0.8,0.8}
	\definecolor{grayyy}{rgb}{0.7,0.7,0.7}
	\definecolor{grayy}{rgb}{0.6,0.6,0.6}
	\begin{figure}
		\vspace{0.1cm}
	\begin{tikzpicture}
		\begin{axis}[name=plot1, y label style={at={(axis description cs:0.15, 0.5), anchor=south}}, xlabel={$k$}, ylabel={$P\of{\text{fault alarm}}$}, xmin=0, ymin=0, xmax=300, ymax=1.05 ,legend style={at={(0.98,0.02)},anchor=south east ,draw=black,fill=white,align=left, nodes={scale=0.7, transform shape}}, height=0.2\textwidth, width=0.241\textwidth]
			\addplot [color=blue, only marks, mark=o] table [x index = {0}, y index = {1},col sep=semicolon]  {plotData/expectedBehaviourStatic.csv};
			\addlegendentry{sFD};
			\addplot [color=red, only marks, mark=x] table [x index = {0}, y index = {1},col sep=semicolon]  {plotData/expectedBehaviourDynamic.csv};
			\addlegendentry{dFD};
			
			\draw[black, line width=0.8mm] (axis cs: 0,0) -- (axis cs: 100,0) -- (axis cs: 100, 1) -- (axis cs: 300, 1);
		\end{axis}
		\begin{axis}[name=fpp,
		at=(plot1.right of north east), anchor=left of north west, xlabel={$k$}, xmin=0, ymin=0, xmax=300, ymax=0.1 ,legend style={at={(0.02,0.98)},anchor=north west ,draw=black,fill=white,align=left, nodes={scale=0.7, transform shape}}, height=0.2\textwidth, width=0.24\textwidth]
		\addplot [color=blue, only marks, mark=o] table [x index = {0}, y index = {1},col sep=semicolon]  {plotData/expectedBehaviourStatic2.csv};
		\addlegendentry{sFD};
		\addplot [color=red, only marks, mark=x] table [x index = {0}, y index = {1},col sep=semicolon]  {plotData/expectedBehaviourDynamic2.csv};
		\addlegendentry{dFD};
		
	\end{axis}
\coordinate[](coord1) at(plot1.below south);
\coordinate[right=2.2cm of coord1](coord2);
		\begin{axis}[name=states, 
			at=(coord2), anchor=north,
			ylabel={$x\uli{3}$ / \SI{}{\meter\per\second}}, xlabel={$k$}, xmin=0, xmax=299, ymax = 8.0, ymin=-8.0, legend style={at={(0.98,0.5)},anchor=south east ,draw=black,fill=white,align=left, nodes={scale=0.7, transform shape}}, height=0.2\textwidth, width=0.45\textwidth]
	
	\addplot [color=black, line width=0.1, mark=none, name path=plus] table [x index = {0}, y index = {1},col sep=semicolon]  {plotData/StatesSimulationSTDPlus3.csv};
	\addplot [color=black, line width=0.1, mark=none, name path=minus] table [x index = {0}, y index = {1},col sep=semicolon]  {plotData/StatesSimulationSTDMinus3.csv};
	\addplot[grayyyy] fill between[of=plus and minus];
	
	\addplot [color=black, line width=0.1, mark=none, name path=plus] table [x index = {0}, y index = {1},col sep=semicolon]  {plotData/StatesSimulationSTDPlus2.csv};
	\addplot [color=black, line width=0.1, mark=none, name path=minus] table [x index = {0}, y index = {1},col sep=semicolon]  {plotData/StatesSimulationSTDMinus2.csv};
	\addplot[grayyy] fill between[of=plus and minus];
	
	\addplot [color=black, line width=0.1, mark=none, name path=plus] table [x index = {0}, y index = {1},col sep=semicolon]  {plotData/StatesSimulationSTDPlus.csv};
	\addplot [color=black, line width=0.1, mark=none, name path=minus] table [x index = {0}, y index = {1},col sep=semicolon]  {plotData/StatesSimulationSTDMinus.csv};
	\addplot[grayy] fill between[of=plus and minus];
\end{axis}
	\end{tikzpicture}
	\caption{Cart-pole simulation results for \num{10000} experiments. \capt{Both upper plots show the probability of fault alarm for each timestep generated by a Monte Carlo simulation. In the left plot, the agent monitored and three other agents have a fault at $k\geq100$ (first scenario). The right plot shows a comparison of the false positive probability for sFD and dFD when the bandwidth is lowered at $k=100$ (second scenario). The gray areas in the lower plot show the 1-, 2- and 3-times standard deviation boundaries of the third state for the first scenario in the case where no countermeasures are taken after fault detection.
		}}
	\label{fig:simResults}
\end{figure}
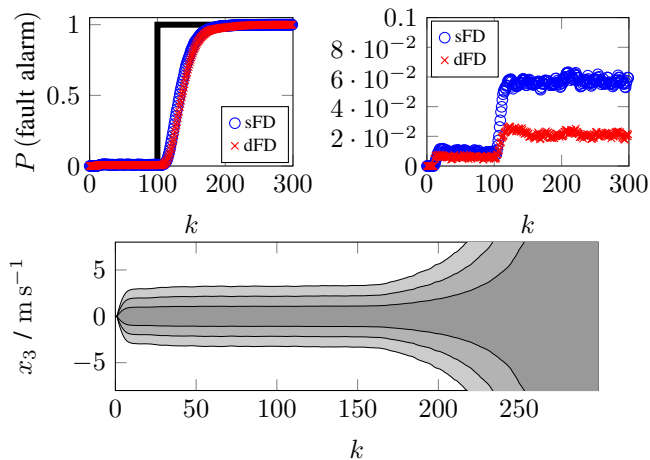

\newcommand{\errorstart}{152}
\newcommand{\errorend}{300}

\newcommand{\errorstartTwo}{300}
\newcommand{\errorendTwo}{300}
\newcommand{\xmaxS}{-8.0}

We simulate 20 inverse cart-pole systems with the same linear dynamics as in \cite{MASTRANGELO2019}.\footnote{The code of the simulation can be found in \url{https://github.com/Data-Science-in-Mechanical-Engineering/PriorityBasedFaultDetection}.}
	The process noise is set to $\matV{i}=3\cdot10^{-4} \mat{I}\ul{4}$, With $\mat{I}\ul{4}$ the $4\times 4$ identity matrix.
	The systems are synchronized using the LQ-approach described in \cite{PredictiveTriggering}, where we select $\matQ{1}=\diag[1, 1, 0, 0]$, $\matQ{2}=\diag[1000, 0, 0, 0]$, and $\ma{R}\uli{i} = 0.1$. Per round, two agents are allowed to send their states.
We use quadratic priorities similar to \cite{Mamduhi_2017} and \cite{Martinez2015}:
\begin{equation*}
	\label{eq:quadrPrio}
	\prio{i}{k}= 	\errorest{ii}{k}{}^\transp\left((\mat{A}-\mat{B}\mat{F}\uli{ii})^\transp\right)^2(\mat{A}-\mat{B}\mat{F}\uli{ii})^2\errorest{ii}{k},
\end{equation*}
with $\mat{A}$ the state, $\mat{B}$ the input, and $\mat{F}\uli{ii}$ the feedback matrix described in \cite{PredictiveTriggering}. Due to the network-induced delays, the priorities calculated at time $k$ determine which messages arrive at $k+2$. The triggering takes this into account \cite{Baumann_2019, MASTRANGELO2019,PredictiveTriggering} and predicts the error two timesteps into the future.

To implement the FD on a real system, priorities have to be quantized. To capture the effect of this quantization, the simulation uses 8-bit values of the priority. We evaluate two scenarios and, for each, we simulate the mult-agent system \num{10000} times. The states and errors are reset to zero at the beginning of each run. Both FD algorithms use $\eta=0.01$, $d=10$, and $b=40$. 

In the first scenario, at time $k=100$, we set the input matrix of the agents $2$--$5$ to zero to simulate the failure of their actors.
 Without a reliable FD, the other agents would not notice this change in behavior and would try to synchronize with the now unstable agents, rendering themselves unstable. In the second scenario, the multi-agent system has no faults, but after time $k=100$, we allow only one agent to communicate, simulating a loss of bandwidth. 

Figure \ref{fig:simResults} shows the results. For every timestep in both top plots, we summed the number of simulation runs that triggered a fault alarm at this timestep and divided that sum by the total number of simulation runs (\num{10000}). This provides an estimate of the probability that the fault detector triggers an alarm at this particular timestep. In the first scenario, before the fault occurrence at $k<100$, both algorithms have a low fault alarm probability (top left plot in Figure (\ref{fig:simResults})). In particular, it is below $\eta=\SI{1}{\percent}$. After the fault at $k=100$, both methods need some time before the fault is detected. The dFD has a lower false positive probability and also detects the error later. This may be a result of the conservativeness of this algorithm. However, as the bottom plot in Figure \ref{fig:simResults} shows, the delay between fault occurrence and detection is sufficiently small for faults to be detected before the system travels too far away.

When the available network bandwidth in the second scenario (top right plot in Figure \ref{fig:simResults}) is lowered, the priorities become larger. As sFD cannot adapt to this, its false positive probability jumps to \SI{5.44}{\percent} (from \SI{0.91}{\percent}). By contrast, dFD changes its decision threshold, and thus its false positive probability only rises to \SI{2.09}{\percent} (from \SI{0.61}{\percent}). Note that changing the network bandwidth alters the distribution of the priority, and therefore Theorem \ref{th:firstorderboundVtwo} no longer holds. 

\subsection{Wireless cart-pole system}
We next implement the algorithm on the testbed described in \cite{PredictiveTriggering, CPSTestbed}.
This testbed contains 20 agents connected over a wireless multi-hop network using Bluetooth Low Energy. Six of these agents are real cart-pole systems and fourteen are simulated using a linear model of a cart-pole system. The goal of each agent is to balance its pole while synchronizing its movement with the other agents. All agents have a microcontroller with 48\,Mhz core frequency.
The communication rounds are executed with a frequency of $10$\,Hz. The FD algorithm is synchronized with the communication rounds. To emulate a fault, we wiggle the top of one pole.

Figure \ref{fig:realsystem} shows the result of one run. When the agent's cart position changes from a slow to a fast and noisy oscillation, the pendulum is shaken. Because the agent changes its behavior during these phases, its priority rises.
The fault detectors detect the fault almost immediately. At the time when the pendulum is shaken, the FD constantly outputs \qq{Fault}. This behavior arises from the summation in (\ref{eq:faultdetectioncond}) for the sFD and the OR operation (at least one period $\mathcal{G}\uli{i, h}$) for the dFD. The time points at which the sFD and dFD trigger a fault alarm are very close. This may be because the priorities jump directly from low to very high values.

The computation footprint of both algorithms is very low. We found the sFD to take approximately $\SI{8}{\micro\second}$ and the dFD about $\SI{23}{\micro\second}$ per agent and timestep on an STM32L476RG ($\SI{80}{\mega\hertz}$ ARM Cortex M4, $\SI{128}{\kilo B}$ SRAM).


\renewcommand{\errorstart}{90}
\renewcommand{\errorend}{124}

\renewcommand{\errorstartTwo}{193}
\renewcommand{\errorendTwo}{222}
\begin{figure}
	\vspace{0.1cm}
	\definecolor{error}{rgb}{1,0.7,0.7}
	\definecolor{normal}{rgb}{0.7,1.0,0.7}
	\pgfdeclarelayer{bg}    
	\pgfsetlayers{bg,main}
	\begin{tikzpicture}
		\begin{axis}[name=states, ylabel={$x\uli{1} / m$}, xmin=0, xmax=240, ymax = 0.2, ymin=-0.2, legend style={at={(0.98,0.5)},anchor=south east ,draw=black,fill=white,align=left, nodes={scale=0.7, transform shape}}, height=0.2\textwidth, width=0.45\textwidth, restrict y to domain=-0.2:0.2]
			\begin{pgfonlayer}{bg}
				\draw [fill=normal, draw=black, ] (axis cs: 0,0.2) rectangle (axis cs: \errorstart,-0.2);
				\draw [fill=error, draw=black, ] (axis cs: \errorstart,0.2) rectangle (axis cs: \errorend,-0.2);
				\draw [fill=normal, draw=black, ] (axis cs: \errorend,0.2) rectangle (axis cs: \errorstartTwo ,-0.2);
				\draw [fill=error, draw=black, ] (axis cs: \errorstartTwo,0.2) rectangle (axis cs: \errorendTwo,-0.2);
				\draw [fill=normal, draw=black, ] (axis cs: \errorendTwo,0.2) rectangle (axis cs: 240,-0.2);
			\end{pgfonlayer}
			\node at (axis cs: 108,0.15){Fault};
			\node at (axis cs: 35,0.15){No Fault};
			\addplot [color=blue, mark=none] table [x index = {0}, y index = {1},col sep=semicolon]  {plotData/RealSystemStates.csv};
		\end{axis}
		
		\begin{axis}[%
			name=kappas,
			at=(states.below south), anchor=north, height=0.15\textwidth, width=0.45\textwidth, xmin=0, xmax=240, ymin=-0.1,ymax=1.1,restrict y to domain=-0.1:1.1, legend style={at={(0.02,0.98)},anchor=north west,draw=black,fill=white, fill opacity=0.6, draw opacity=1.0, text opacity=1.0, align=left, nodes={scale=0.7, transform shape}}, ytick={0,1}, yticklabels={No Fault, Fault}]
			\addplot [color=blue, mark=none] table [x index = {0}, y index = {1},col sep=semicolon]  {plotData/RealSystemFaultDetectedSum.csv};
			\addlegendentry{sFD};
			\addplot [color=red, mark=none] table [x index = {0}, y index = {1},col sep=semicolon]  {plotData/RealSystemFaultDetected.csv};
			\addlegendentry{dFD};
			
		\end{axis}
	\end{tikzpicture}
	\caption{Experimental run on wireless cart-pole system. \capt{The plots show: the first state (position of the cart) (top) and the output of the FD (bottom); 
			The green and red areas in the top plot mark the times where the fault detector with time-variant threshold outputs \qq{No Fault}/\qq{Fault}.}}
	\label{fig:realsystem}
\end{figure}
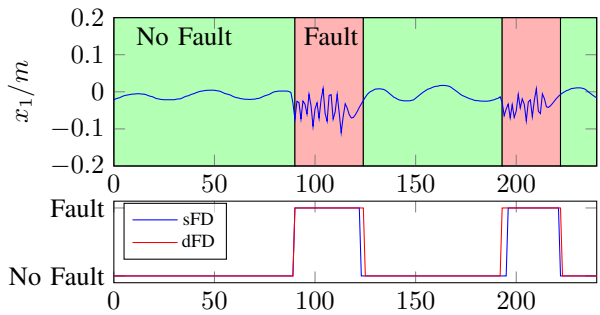
	\section{Conclusion and Outlook}
Reliable remote fault detection is important to increase the safety of multi-agent systems. In this work, we investigate fault detection for a multi-agent system controlled using a recently proposed event-triggering structure called predictive triggering. Instead of spare state measurements transmitted by an event-triggered communication protocol, we propose using low-dimensional communication priorities to dependably detect faults.
Our methods use a decision threshold. While this threshold remains constant for the first method (sFD), the second method (dFD) adapts it according to the communication history. We optimized the computational cost of our methods and showed that they work in real-time on a small microcontroller. Both methods have been shown to detect faults. sFD requires less memory than dFD while exhibiting similar detection behavior. However, dFD is less likely to have false positives in case of changing bandwidth.

In cases, where a fault does not change the magnitude of the priorites, the presented FDs do not detect a fault. Here, other methods should be used.

We have used MC to approximate the analytically intractable thresholds directly. In order to generate thresholds with strict guarantees, one could use concentration inequalities instead. However, these thresholds will typically be more conservative. Thus, the results of this paper suggest using MC as a suitable alternative.


	\bibliographystyle{IEEEtran}
	\bibliography{references} 

\begin{thebibliography}{10}
\providecommand{\url}[1]{#1}
\csname url@samestyle\endcsname
\providecommand{\newblock}{\relax}
\providecommand{\bibinfo}[2]{#2}
\providecommand{\BIBentrySTDinterwordspacing}{\spaceskip=0pt\relax}
\providecommand{\BIBentryALTinterwordstretchfactor}{4}
\providecommand{\BIBentryALTinterwordspacing}{\spaceskip=\fontdimen2\font plus
\BIBentryALTinterwordstretchfactor\fontdimen3\font minus
  \fontdimen4\font\relax}
\providecommand{\BIBforeignlanguage}[2]{{%
\expandafter\ifx\csname l@#1\endcsname\relax
\typeout{** WARNING: IEEEtran.bst: No hyphenation pattern has been}%
\typeout{** loaded for the language `#1'. Using the pattern for}%
\typeout{** the default language instead.}%
\else
\language=\csname l@#1\endcsname
\fi
#2}}
\providecommand{\BIBdecl}{\relax}
\BIBdecl

\bibitem{Baumann_2021}
D.~Baumann, F.~Mager, U.~Wetzker, L.~Thiele, M.~Zimmerling, and S.~Trimpe,
  ``Wireless control for smart manufacturing: Recent approaches and open
  challenges,'' \emph{Proceedings of the IEEE}, 2021.

\bibitem{Yaqoob2020}
I.~Yaqoob, L.~U. Khan, S.~M.~A. Kazmi, M.~Imran, N.~Guizani, and C.~S. Hong,
  ``Autonomous driving cars in smart cities: Recent advances, requirements, and
  challenges,'' \emph{IEEE Network}, 2020.

\bibitem{Isermann_2005}
R.~Isermann, \emph{Fault-diagnosis systems: an introduction from fault
  detection to fault tolerance}.\hskip 1em plus 0.5em minus 0.4em\relax
  Springer, 2005.

\bibitem{NCSET}
W.~Heemels, K.~Johansson, and P.~Tabuada, ``An introduction to event-triggered
  and self-triggered control,'' in \emph{51st IEEE Conference on Decision and
  Control}, 2012.

\bibitem{Baumann_2019}
S.~Trimpe and D.~Baumann, ``{Resource-Aware IoT Control: Saving Communication
  Through Predictive Triggering},'' \emph{IEEE Internet of Things Journal},
  2019.

\bibitem{MASTRANGELO2019}
J.~M. Mastrangelo, D.~Baumann, and S.~Trimpe, ``Predictive triggering for
  distributed control of resource constrained multi-agent systems,''
  \emph{IFAC-PapersOnLine}, 2019, 8th Workshop on Distributed Estimation and
  Control in Networked Systems NECSYS.

\bibitem{PredictiveTriggering}
F.~Mager, D.~Baumann, C.~Herrmann, S.~Trimpe, and M.~Zimmerling, ``Scaling
  beyond bandwidth limitations: Wireless control with stability guarantees
  under overload,'' \emph{ACM Trans. Cyber-Phys. Syst.}, 2021.

\bibitem{WILHELM2021}
Y.~Wilhelm, P.~Reimann, W.~Gauchel, and B.~Mitschang, ``Overview on hybrid
  approaches to fault detection and diagnosis: Combining data-driven,
  physics-based and knowledge-based models,'' \emph{Procedia CIRP}, 2021.

\bibitem{Zhang_2004}
P.~Zhang, S.~Ding, P.~Frank, and M.~Sader, ``Fault detection of networked
  control systems with missing measurements,'' in \emph{5th Asian Control
  Conference}, 2004.

\bibitem{Zhang2007}
P.~Zhang and S.~X. Ding, ``Fault detection of networked control systems with
  limited communication,'' in \emph{Fault Detection, Supervision and Safety of
  Technical Processes 2006}.\hskip 1em plus 0.5em minus 0.4em\relax Elsevier
  Science Ltd, 2007.

\bibitem{Wang_2009}
Y.~Wang, H.~Ye, S.~X. Ding, G.~Wang, and D.~Zhou, ``Residual generation and
  evaluation of networked control systems subject to random packet dropout,''
  \emph{Automatica}, 2009.

\bibitem{Wang_2016}
Y.-L. Wang, P.~Shi, C.-C. Lim, and Y.~Liu, ``Event-triggered fault detection
  filter design for a continuous-time networked control system,'' \emph{IEEE
  Transactions on Cybernetics}, 2016.

\bibitem{Wang_2017}
Y.-L. Wang, C.-C. Lim, and P.~Shi, ``Adaptively adjusted event-triggering
  mechanism on fault detection for networked control systems,'' \emph{IEEE
  Trans. on Cybernetics}, 2017.

\bibitem{Solowjow2_2018}
F.~Solowjow, D.~Baumann, J.~Garcke, and S.~Trimpe, ``Event-triggered learning
  for resource-efficient networked control,'' in \emph{Annual American Control
  Conference}, 2018.

\bibitem{NetworkControlGoesWireless}
F.~Mager, D.~Baumann, R.~Jacob, L.~Thiele, S.~Trimpe, and M.~Zimmerling,
  ``Feedback control goes wireless: Guaranteed stability over low-power
  multi-hop networks,'' in \emph{10th ACM/IEEE International Conference on
  Cyber-Physical Systems}, 2019.

\bibitem{Mixer}
C.~Herrmann, F.~Mager, and M.~Zimmerling, ``Mixer: Efficient many-to-all
  broadcast in dynamic wireless mesh networks,'' in \emph{16th ACM Conference
  on Embedded Networked Sensor Systems}, 2018.

\bibitem{CPSTestbed}
D.~Baumann, F.~Mager, H.~Singh, M.~Zimmerling, and S.~Trimpe, ``Evaluating
  low-power wireless cyber-physical systems,'' in \emph{IEEE Workshop on
  Benchmarking Cyber-Physical Networks and Systems}, 2018.

\bibitem{Mamduhi_2017}
M.~H. Mamduhi, A.~Molin, D.~Tolić, and S.~Hirche, ``Error-dependent data
  scheduling in resource-aware multi-loop networked control systems,''
  \emph{Automatica}, 2017.

\bibitem{Martinez2015}
M.~Mart\'{i}nez-Rey, F.~Espinosa, A.~Gardel, C.~Santos, and A.~Salcedo,
  ``Computation reduction of smart sensors for ebse using mahalanobis
  sampling,'' in \emph{International Conference on Event-based Control,
  Communication, and Signal Processing}, 2015.

\end{thebibliography}
	\appendices
	
\end{document}